# Single-Phase Duodenary High-Entropy Fluorite/Pyrochlore Oxides with an Order-Disorder Transition

Andrew J. Wright [a], Qingyang Wang [b], Chongze Hu [c], Yi-Ting Yeh [a], Renkun Chen [b,c], Jian Luo [a,c,*]

[a] Department of NanoEngineering; [b] Department of Mechanical & Aerospace Engineering; [c] Program of Materials Science and Engineering, University of California, San Diego, La Jolla, CA 92093, USA

## Abstract

Improved thermomechanical properties have been reported for various high-entropy oxides containing typically five metal cations. This study further investigates a series of duodenary (11 metals + oxygen) high-entropy oxides by mixing different fractions of a five-cation fluorite-structured niobate and a seven-cation pyrochlore (both containing Yb) with matching lattice constants. Nine compositions of duodenary high-entropy oxides have been examined. All of them exhibit single high-entropy phases of either disordered fluorite or ordered pyrochlore structure. An order-disorder transition (ODT) is evident with changing composition, accompanied by a reduction in thermal conductivity ($k$). In comparison with the ODT criteria developed from ternary oxides, these duodenary oxides are more prone to disorder, but the ODT is still controlled by similar factors (but at different thresholds). Interestingly, there are abrupt increases in Young's modulus ($E$) at low mixing concentrations near both endmembers. The $E/k$ ratios are increased, in comparison with both endmembers. This study suggests a new route to tailor high-entropy ceramics via controlling cation ordering *vs.* disordering.

*Correspondence should be addressed to J.L. (email: jluo@alum.mit.edu)

# Graphical Abstract

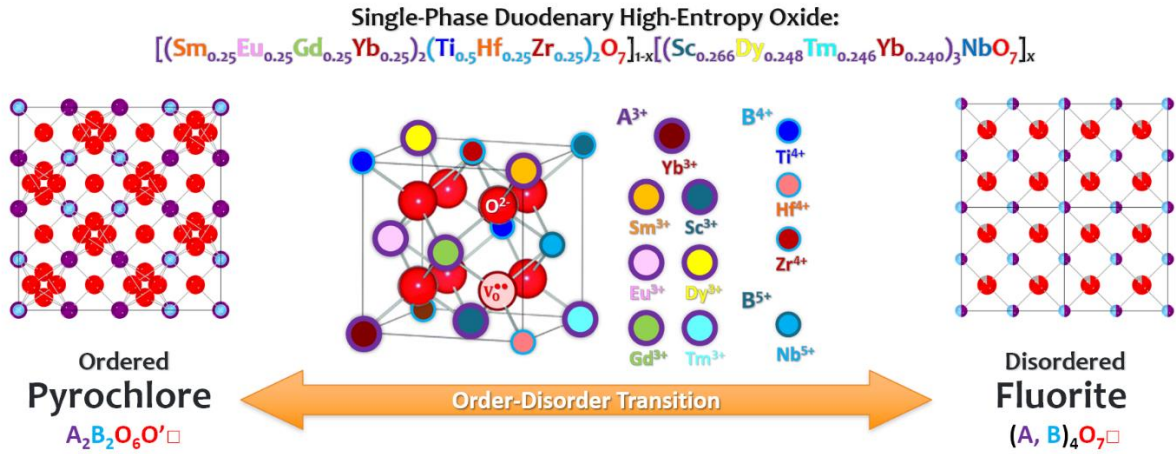

**Highlights:**

- A series of single-phase duodenary high-entropy oxides are synthesized.
- An order-disorder transition occurs with changing composition.
- Increased modulus-to-thermal conductivity ratios from the rule of mixture.
- Unexpected increases in Young's modulus at low doping levels.
- Temperature-dependence reveals ultralow glass-like thermal conductivity.

# 1 Introduction

High-entropy ceramics (HECs), including rocksalt [1], perovskite [2], fluorite [3], and pyrochlore [4,5] oxides, borides [6,7], carbides [8–11], and silicides [12,13] have attracted great research interests. Recently, it is further proposed to broaden HECs to compositionally-complex ceramics (CCCs) [14–16]. The majority of the past studies focused on HECs with about five (sometimes four, and, in a few cases, six) cations. Here, the only few exceptions are represented by high-entropy $ABO_3$ perovskite $(Gd_{1/5}La_{1/5}Nd_{1/5}Sm_{1/5}Y_{1/5})(Co_{1/5}Cr_{1/5}Fe_{1/5}Mn_{1/5}Ni_{1/5})O_3$ [17] and high-entropy $A_2B_2O_7$ pyrochlore $(La_{1/7}Ce_{1/7}Pr_{1/7}Nd_{1/7}Sm_{1/7}Eu_{1/7}Gd_{1/7})_2(Sn_{1/3}Hf_{1/3}Zr_{1/3})_2O_7$ [18], where 10 metal cations are distributed over two cation sublattices. However, each of these two studies only included one 10-cation composition, and both cases are ordered phases only (so there are 3-7 cations in each sublattices). Notably, no systematic study has been reported to investigate how the change in composition affects the phase stability (particularly cation order and disorder) and subsequently the properties of such many-cation HECs.

One distinct property of HECs and CCCs is represented by their reduced thermal conductivities that have been observed across nearly all systems examined [3–5,12,15,16,19–31]. This is presumably due to the increased disorder in mass, size, and valency in solid solutions with five or more different cations [16,27,32–35]. From the materials science point of view, reduced thermal conductivities typically come at the expense of Young's modulus [14,27]. For HECs, however, the modulus appears to be retained while the thermal conductivity is remarkably reduced [12,15,16,22,27,28], thereby being scientifically interesting and technologically useful.

Fluorite and pyrochlore oxides demonstrate the ultralow thermal conductivities ($k$'s) with values approaching 1 W·m$^{-1}$·K$^{-1}$ [16,36–38]. With their Young's moduli ($E$) of $200 - 250\ GPa$, these oxides exhibit the highest $E/k$ ratios so far [15,16,36]. Various traditional (typically ternary) fluorite or pyrochlore oxides have widely been used or examined as thermal barrier coatings (TBCs) [39–45]. The pyrochlore and fluorite crystal structures have a direct connection to each other through an order-disorder transition (ODT) of the cations and anions [46–51]. Previous studies have examined this ODT through diffraction techniques and measurements of ionic and thermal conductivities [44,52–58]. However, all these prior studies were conducted on simpler ternary (and occasionally quaternary) oxides. The ODT in HECs, which has the potential for opening a new window to tailor thermomechanical properties, has not been examined.

In this study, we first synthesized a series of duodenary (11 metals + oxygen) oxides by mixing a five-cation disordered fluorite niobate and a seven-cation ordered pyrochlore oxide (both containing Yb) with negligible lattice constant and density mismatch. Here, we use "duodenary oxides" to refer to oxides with 11 unique metal cations (12 elements including oxygen), similar to the terminology in literature: "binary (or ternary) oxides," referring to oxides with one (or two) metal cation(s) plus oxygen. The focus and novelty of this study lies in the first investigation of an ODT in HECs with changing composition and its influence on thermomechanical properties, along with the first investigation of a series of duodenary HECs with systematic compositional variation. An abnormal increase in Young's modulus above the rule of mixture (RoM) average is observed, and it abruptly occurs at low doping levels. Consequently, the *E/k* ratio also increases above the RoM averages of the two endmembers.

## 2    Pyrochlore *vs.* Fluorite and Order-Disorder Transition (ODT)

Figure 1 highlights the differences between the pyrochlore and fluorite phases and three conventional routes to induce the ODT. The ideal pyrochlore (space group 227, $Fd\bar{3}m$) is an ordered (cations and anions) 2 × 2 × 2 variant of the disordered fluorite unit cell. This ordered $A_2B_2O_7$ pyrochlore structure, where the $AO_8$ polyhedron is a distorted scalenohedron and the $BO_6$ polyhedron is a perfect octahedron, is signified by the 48*f* oxygen positional parameter $x = 0.3125$. Disordering initiates as the 48*f* site migrates towards $x = 0.375$ (or 3/8, the ideal oxygen position in a disordered fluorite phase [48]). It is energetically unfavorable for the 48*f* oxygen in the pyrochlore phase to have $x \gtrsim 0.36$ due to an increasing O-O repulsive force. Instead, the 48*f* oxygen begin to populate the 8*a* vacancy sites, thus initiating the disordering process [53]. In the schematic of the fluorite structure, the $AO_8$ polyhedron approaches a perfect cube while the $BO_6$ polyhedron becomes a distorted octahedron. However, the perfect polyhedrons are not reached in the real disordered fluorite structure as the cations disorder leads to the formation of (A, B)$O_7$ polyhedrons on average (instead of the perfect $AO_8$ and $BO_6$ polyhedrons).

The three most common routes to induce the ODT are via increasing temperature, the difference in the cation radii, and nonstoichiometry [59]. First, most rare earth zirconate and hafnate pyrochlores undergo temperature-induced disordering transition at $\geq 1600°C$ [60]. Second, the ordered pyrochlore structure is preferred if the size difference between the A- and B-site cations is significant, while the disordered fluorite structure is preferred if the cations are of similar radii. Here, the critical radius ratio for the ODT is approximately 1.46 (disordering if $r_A^{VIII}/r_B^{VI} < \sim 1.46$), where the radii of the A- and B-site are assumed to be eight- and six-coordinated [50,59]. Third, non-stoichiometry arising from an imbalance in the concentration of A- and B-site cations, of the form $A_{2\pm y}B_{2\pm y}O_{7\pm \delta}$, can induce an ODT. For rare earth zirconates and hafnates, the ODT (disordering) occurs when $y > 0.1 - 0.2$ [60].

## 3 Experimental Procedures

### 3.1 Design of Compositions

On one side, we selected $[(Sm_{0.25}Eu_{0.25}Gd_{0.25}Yb_{0.25})_2(Ti_{0.5}Hf_{0.25}Zr_{0.25})_2O_7]_{1-x}$ (denoted as P1) and its two variations (with different atomic fractions on the B site, denoted as P1' and P1") in the ordered pyrochlore structure that we have investigated recently [16] as one (ordered pyrochlore) endmember. They are the high-entropy version of the rare earth (RE)-stabilized zirconia $(RE^{3+})_2(Zr^{4+})_2O_7$ or hafnia $(RE^{3+})_2(Hf^{4+})_2O_7$ that are commonly used in TBCs.

On the other side, we made fluorite-structured $[(Sc_{0.266}Dy_{0.248}Tm_{0.246}Yb_{0.240})_3NbO_7]_x$ (N1) $[(Sc_{0.333}Dy_{0.226}Tm_{0.224}Yb_{0.218})_3NbO_7]_{0.5}$ (N1'), and $[(Y_{0.266}Dy_{0.248}Tm_{0.246}Yb_{0.240})_3NbO_7]_{0.5}$ (N2) as the other (disordered fluorite) endmember. They are the high-entropy version of rare earth niobates $(RE^{3+})_3(Nb^{5+})O_7$ in the disordered fluorite structure (space group 225, $Fm\overline{3}m$), where both the cations and 1/8 oxygen vacancies per unit cell are disordered. These rare earth niobates have ultralow thermal conductivities [36–38]. Noting that the specific compositions were designed so there are matches in both lattice parameters and densities of the two endmembers.

We then fabricated duodenary high-entropy oxides by mixing three pairs of endmembers. In the P1N1 series, we mixed P1: $[(Sm_{0.25}Eu_{0.25}Gd_{0.25}Yb_{0.25})_2(Ti_{0.5}Hf_{0.25}Zr_{0.25})_2O_7]_{1-x}$ and N1: $[(Sc_{0.266}Dy_{0.248}Tm_{0.246}Yb_{0.240})_3NbO_7]_x$ in the volumetric ratios of 100:0, 98:2, 90:10, 75:25, 50:50, 25:75, 10:90, 2:98, and 0:100, respectively. Here, for example, P1N1 50:50 refers the composition mixed (P1)$_{50}$(N1)$_{50}$ (but it is a single-phase pyrochlore after the mixing). We also made two extra compositions, P1'N1' 50:50 [= (P1')$_{50}$(N1')$_{50}$] and P1"N2 50:50 [= (P1")$_{50}$(N2)$_{50}$], where P1', P1" and N1' are derivative compositions of P1 and N1 with varying atomic ratios, and N2 is another high-entropy niobate of a different cation combination (with Y to replace Sc in N1/N1'). The compositions and nomenclatures of all specimens studied are shown in Table 1.

It is important to emphasize that all nine mixed compositions are single-phase (either fluorite- or pyrochlore-structured) high-entropy solid solutions after mixing.

### 3.2 Materials and Synthesis

The specimens were consolidated by first synthesizing the pure phase of the high-entropy pyrochlore (P1, P1', and P1") and high-entropy niobate (N1, N1' and N2) endmembers in parallel. For each endmember, stoichiometric amounts of the constituent binary oxides with particle sizes of ~5 μm (purchased from US Research Nanomaterials) were weighed with 0.01 mg precision (Secura125-1S, Sartorius, Germany). A total of 15 g was weighed for the desired composition and placed in a 100 ml $Y_2O_3$-stabilized $ZrO_2$ (YSZ) planetary mill jar. YSZ grinding media was added to the jar in a ball-to-powder mass ratio of 10:1. Ten milliliters of ethanol was also added to the jar. The jar was planetary milled at 300 RPM for 24 h (PQN04, Across International, United States). The contents were transferred to a 1 L glass beaker and dried at 85°C overnight. The dried powder was collected, pressed in a 1" stainless steel die, and calcined at 1600°C for 12 h. These specimens of endmembers were then coarsely ground back into a powder with an agate mortar and pestle.

The coarse single-phase powders of two endmembers were then mixed in the desired volume ratios for a total weight of 2 g. The mixed powders were placed in a poly(methyl methacrylate) high-energy ball mill (HEBM) vial with tungsten carbide (WC) inserts and one Ø5/16" WC ball. Additionally, 2 wt% stearic acid was added to the vial to serve as a lubricant and binder. The vials were dry-milled for 100 minutes (SPEX 8000D, SPEX SamplePrep, USA). Following milling, the powders were uniaxially pressed in a Ø13 mm stainless steel die and placed on Pt foil in an $Al_2O_3$ crucible to undergo sintering in air at 1600°C for 24 h. Lastly, the sintered pellets were ground with a 30 μm diamond disc before characterization.

### 3.3 Characterization

#### 3.3.1 X-ray Diffraction (XRD) and Density ($\rho$)

A Miniflex II XRD (Rigaku, Japan) operating at 30 kV and 15 mA was used to collect data on specimens at 0.02° 2θ steps for 2θ of 20 – 90° with a 2 s dwell time per step. Rietveld refinements through GSAS-II [61] were used to obtain the lattice constant and theoretical density.

The bulk density ($\rho$) of each specimen was determined through the boiling method abiding by the ASTM Standard C373-18 [62]. The relative density was computed by dividing the measured bulk density by the theoretical density. All specimens were found to have relative densities within $96 - 99$ %, except for P1"N2 50-50 with a lower relative density of 91.3 %.

### 3.3.2 Scanning Electron Microscopy (SEM)

Specimens were hot-mounted in acrylic and polished to 40 nm colloidal silica. A scanning electron microscope (SEM, FEI Apreo, OR, USA) operating at 20 kV was used to examine the microstructure. Elemental maps were generated through energy dispersive spectroscopy (EDS, Oxford N-Max[N]) to probe the compositional homogeneity.

### 3.3.3 Scanning Transmission Electron Microscopy (STEM)

Scanning transmission electron microscopy (STEM) specimens were prepared by a dual-beam focused ion beam (FIB, FEI Scios DualBeam, OR, USA). A 300 kV, double aberration-corrected STEM (JEOL JEM-ARM300CF, Japan) microscope was used to generate atomic resolution images with a high-angle annular dark-field (HAADF) detector of an inner collection angle ~ 80 mrad. Atomic resolution EDS elemental maps were collected over five minutes with a probe current of ~ 300 pA. The STEM-HAADF image was passed through the default bandpass filter available through Gatan Microscopy Suite 3.3 (GMS 3.3, Gatan, CA, USA). A radial Wiener filter was used to reduce the noise of the EDS elemental maps. The GMS 3.3 plug-in was available in a free software package offered by HREM Research Inc..

### 3.3.4 Young's Modulus (*E*)

The Young's modulus of the specimens was determined through a pulse-echo sonic resonance setup following the ASTM Standard C1198-20 [63]. The longitudinal ($u_L$) and transverse ($u_T$) wave speeds were determined for each of the specimens. Poisson's ratio ($v$) and Young's modulus were determined from Eqs. (1) and (2) below:

$$v = \frac{u_L^2 - 2u_T^2}{2(u_L^2 - u_T^2)} \qquad (1)$$

$$E_{measured} = 2u_T^2 \rho (1 + v) \qquad (2)$$

The modulus was corrected for porosity [64] through Eq. (3):

$$E = \frac{E_{measured}}{1 - 2.9P} \tag{3}$$

Here, $P$ is the fraction of pores.

### 3.3.5 Thermal Conductivity ($k$)

The thermal diffusivity ($\alpha$) was measured at 25°C and then from 200°C – 1000°C in 200°C steps through laser flash analysis (LFA 467 *HT HyperFlash*, NETZSCH, Germany). Before measurement, the specimens had a carbon coating sprayed onto the top and bottom surfaces to maximize laser absorption and infrared emission. The values obtained from $T > 200°C$ were corrected for external radiation. However, internal radiation is still present as it is inherent to the specimen [65]. The Neumann-Kopp rule was used to estimate the specific heat capacity ($c_p$) using tabulated values of the constituent binary oxides [66]. The thermal conductivity ($k$) was determined through the product of thermal diffusivity, density, and specific heat capacity, as shown in Eq. (4).

$$k_{measured} = \alpha \rho c_p \tag{4}$$

Similarly to the modulus, the thermal conductivity was corrected for porosity [67] by Eq. (5).

$$k = \frac{k_{measured}}{(1 - P)^{3/2}} \tag{5}$$

## 4 Results and Discussion

### 4.1 Formation of Single High-Entropy Phases and ODT

The pyrochlore endmember P1 (P1N1 100-0) came from a prior publication from our group, $(Sm_{1/4}Eu_{1/4}Gd_{1/4}Yb_{1/4})_2(Ti_{1/2}Hf_{1/4}Zr_{1/4})_2O_7$, which displayed an attractive $E/k$ ratio (175.2 $\frac{GPa \cdot m \cdot K}{W}$) [16]. The N1 (P1N1 0-100) high-entropy niobate, $(Sc_{0.266}Dy_{0.248}Tm_{0.246}Yb_{0.240})_3NbO_7$, was designed so that the lattice parameter was half that of the pyrochlore ($a_{N1} = \frac{1}{2}a_{P1} = 5.163$ Å) and the densities were identical ($\rho_{P1} = \rho_{N1} = 7.37 \; g/cm^3$). This was designed to remove any potential effects from lattice or density mismatch. A calibration curve (Figure S1 in Suppl. Data) was used to determine the elements and exact molar amounts to obtain desired lattice parameters and densities. The pyrochlore calibration curve was generated by data from a prior publication from our group [16] while the niobate calibration curve is unpublished.

The XRD patterns of the P1N1 series are shown in Figure 2(a). Figure 2(b) shows the logarithmic intensity of the peak evolution of the (331) superstructure peak for the ordered pyrochlore phase. All specimens exhibit a single high-entropy solid solution phases (of either pyrochlore or fluorite structure). No visible secondary phase peak was present in any XRD patterns. It is evident that the lattice parameter variation is negligible in this series with changing $x$ in the mixed $P1_{1-x}N1_x$ compositions.

The formation of single high-entropy phase is further supported by the compositional and microstructural homogeneity. For example, Figure 3 shows the microstructure of the polished cross-section along with the elemental maps for all the cations for P1N1 50:50. The backscattered electron image shows no distinct contrast from different mass (Z) beyond slight grain orientation contrast. The grain size is ~10-20 μm. The EDS elemental maps suggest the material is homogenous, albeit slight Zr agglomeration. The overall metal cation composition was quantified and compared to theoretical values calculated through the stoichiometry. Both values are in excellent agreement, and differences are within the EDS errors. Additional EDS elemental maps are shown in Suppl. Figs. 2 and 3, which all suggest homogenous compositions (that agreed with theoretical/nominal values) supporting the formation of single high-entropy phases.

Notably, the pyrochlore structure vanishes (*i.e.*, with an ODT) slightly above $x = 75$ % (*i.e.*, the (331) superstructure peak is barely observed in P1N1 75:25 but disappears in P1N1 90:10). EDS elemental mapping confirmed that the compositions are homogenous in both the pyrochlore (Suppl. Fig. S2 for P1N1 25-75) and fluorite (Suppl. Fig. S3 for P1N1 10-90) phases just before and after the occurrence of the ODT.

### 4.2 Atomic-Resolution Structure and Composition

To further characterize the atomic resolution structure and composition, particularly the distribution of cations like Sc in A *vs.* B sublattice in the ordered $A_2B_2O_7$ pyrochlore phase, a TEM foil of the specimen P1N1 50-50 was prepared by FIB-SEM lift-out. In the STEM, the specimen was tilted to the [110] zone axis because it is the lowest index where A- and B-sites can be separated. The low magnification STEM-HAADF image in Figure 4(a) shows structural homogeneity with no observable nanodomains and clustering.

The higher magnification image in Figure 4(b) reveals the different Z contrasts from different sites in the pyrochlore structure due to larger, higher Z (brighter) atoms on the A-site and smaller, lower Z elements on the B-site. A simulated STEM image was generated from a $10 \times 10 \times 10$ supercell and compared to the raw STEM-HAADF image, which shows a good match of the intensity modulations from the alternating A-site and mixed-site rows.

Figure 4(c) and (d) further show the EDS collection area and the bandpass filtered STEM-HAADF image, which makes the site contrast clearer along with the filtered elemental maps for Yb and Sc as an example for an A- and B-site cation. The elemental maps suggest that Yb atoms are on A-site and Sc are on the B-site, respectively. Note that Sc is the only element with some ambiguity on which site to occupy solely based on the atomic radius. Here the atomic resolution element map directly confirm that Sc atoms sit on B-site.

In summary, the STEM-HAADF-EDS analysis, in conjunction with the SEM-EDS and XRD, unequivocally showed the formation of an ordered single high-entropy pyrochlore phase in P1N1 50:50, with compositional homogeneity at both micrometer and atomic scale. It further clarified that Sc occupies the B-site the ordered $A_2B_2O_7$ pyrochlore.

### 4.3    Sc Influence on Order *vs.* Disorder

Upon adding the high-entropy niobate endmember into the high-entropy pyrochlore, it is well expected that the large 3+ rare earth *f*-elements, such as Dy, Tm, and Yb, prefer the A-site of the pyrochlore structure where Sm, Eu, Gd, and Yb already reside. Niobium ($r_{Nb}^{VI} = 0.64$ Å) can then be assumed to go to the B-site where Ti ($r_{Ti}^{VI} = 0.605$ Å), Zr ($r_{Zr}^{VI} = 0.72$ Å), and Hf ($r_{Hf}^{VI} = 0.71$ Å) [68] are, since these cations are relatively small. The preferred site of Sc is the only one in question. The Sc cation has a 3+ valency that is suited for the A-site, but it is also small enough to fit on the B-site ($r_{Sc}^{VI} = 0.745$ Å). Prior studies (albeit not on high-entropy compositions) by Allpress and Rossell concluded that Sc has little tendency to enter the A-site in the pyrochlore structure [69]. When excess Sc was added to the pyrochlore structure, Sc precipitated out in a second phase before any significant amount entered the A-site. Our atomic resolution STEM-EDS analysis directly confirmed that Sc atoms sit on the B-site.

In addition to its small size, the occupation of Sc on B-site also helps to keep stoichiometric A/B ratio and the mixing of $Sc^{3+}$ and $Nb^{5+}$ cations on the B-site also supplies a charge balance (to the average of 4+ in normal pyrochlore); both will be ideally maintained if $Sc^{3+}$ and $Nb^{5+}$ atomic fractions are 1/4 in the high-entropy niobate endmember. However, there is less $Sc^{3+}$ cations in the N1 endmember. This implies that other, larger, rare earth elements have to fill some of the B-site in the P1N1 series, which would like to destabilize the pyrochlore structure and promote disorder.

To test this hypothesis, two additional fluorite-structured high-entropy niobate endmembers (N1' and N2) were designed and synthesized. The first one modified P1N1 0-100 by increasing the concentration of Sc to 25% of the total cations in the high-entropy niobate endmember N1': $(Sc_{0.333}Dy_{0.226}Tm_{0.224}Yb_{0.218})_3NbO_7$ to enable equiatomic $Sc^{3+}$ and $Nb^{5+}$ cations on the B-site as well as the ideal 1:1 A/B stoichiometric ratio. In the second case, we replaced the Sc with Y to make a Sc-free high-entropy niobate endmember N2: $(Sc_{0.266}Dy_{0.248}Tm_{0.246}Yb_{0.240})_3NbO_7$. Subsequently, both N1' and N2 were mixed in a 50:50 vol. % ratio with designed P1' and P2" (variations of P1 with different cation ratios to ensure negligible lattice and density mismatch).

The XRD results of these two duodenary oxides are shown in Figure 5. As expected, the P1'N1' 50-50 that enables the ideal 1:1 A/B stoichiometric ratio with an equal amount of Sc and Nb in B-site possessed the ordered pyrochlore phase. On the other hand, the Sc-free P1"N2 50-50 become a disordered fluorite structure (presumably because larger rare earth elements would have to occupy at least 25% B-site to destabilize the ordered $A_2B_2O_7$ structure if it were pyrochlore).

### 4.4 Pyrochlore Stability Rules

We may further explore how the ODT rules obtained from simpler ternary and quaternary oxides transform to many-component (such as duodenary) high-entropy oxides. As we have discussed in §2 and schematically shown in Figure 1, the three common routes to induce a pyrochlore to fluorite transformation or ODT are through increasing temperature, size ratio ($r_A/r_B$), and nonstoichiometry ($A/B$).

The temperature effect was not investigated in this study as all of our specimens were sintered at 1600°C. Prior studies showed the lowest ODT temperature is ~1575°C in $Gd_2Zr_2O_7$ [70], and the ODT temperatures are > 1600°C for all other stoichiometric ternary pyrochlores [60].

The $r_A/r_B$ ratios for specimens are shown in Figure 6(a). Prior studies, based mostly on stoichiometric ternary oxides, suggested that the ordered pyrochlore structure should form when $r_A/r_B > 1.46$ with an estimated error of $\pm 0.003$ [71]. Our results of duodenary oxides showed an ODT (or disordering) occurred at higher threshold of $r_A/r_B \sim 1.48$ in the P1N1 series. The observed higher tendency to disorder can be explained from two facts. First, high-entropy solid solutions with severe lattice distortion promote disordering. Second, the prior threshold of ~1.46 was proposed based on stoichiometric ternary oxides. In the P1N1 series, there is Sc deficiency, so that some other larger rare earth elements have to fill a small fraction (<5%) of the B-site, which will also promote disordering. This second proposed cause is further supported by the disordering of Sc-free of P1"N2 50:50 with $r_A/r_B$ of ~1.49, since more larger rare earth elements have to fill (25% of) the B-site, which strongly promotes disordering.

Similarly, the A/B ratios for specimens are shown in Figure 6(b). Here, we estimate the critical A/B ratio from the mean of all available ODTs in relevant ternary oxides, *e.g.*, $Sm_2Ti_2O_7$, $Sm_2Hf_2O_7$, $Sm_2Zr_2O_7$, $Eu_2Ti_2O_7$, at 1600°C (but excluding $Gd_2Zr_2O_7$, $Yb_2Hf_2O_7$, and $Yb_2Zr_2O_7$, which do not possess a pyrochlore phase at 1600°C ). This estimation based on ternary oxides suggests that the disordered fluorite structure should form when the $A/B$ ratio is $> 1.16 \pm 0.04$ (for B-site deficient, with estimated 3% errors). Here, our results of duodenary oxides show that an ODT (or disordering) occurred at a critical $A/B$ ratio between 1.16 and 1.20 in the P1N1 series. Again, high-entropy solid solutions with severe lattice distortion promotes disordering and shift the threshold of the A/B ratio a bit higher (albeit the difference is within the errors). The two additional specimens have much larger (P1"N2) and smaller (P1'N1') A/B ratios, which are consistently disordered and ordered, respectively.

### 4.5 Room Temperature Thermomechanical Properties

Figure 7(a) displays the measured room temperature Young's modulus of all the duodenary oxides. Young's modulus has a noticeable increase in all mixed compositions of P1N1 above the rule of mixtures (RoM) averages. More interestingly, sharp increases occur at low concentrations of mixing at both sides. Such the sharp increases was explained from a strongly negative excess volume of mixing or a negative enthalpy of mixing in a prior study [72]. The negative excess volume of mixing is unlikely in this case because the measured lattice parameters do not deviate significantly from Vegard's law, as shown in Figure 7(b). A recent study suggested that a negative enthalpy of mixing in pyrochlore-fluorite $Ho_2Ti_{2-x}Zr_xO_7$ resulted in short-range ordering [73]. We hypothesize that similar short-range chemical or structural orders, which are almost inevitable for such complex high-entropy oxides according to a recent thermodynamic analysis [74], can explain our observation of increased stiffness in this duodenary P1N1 series. Moreover, a softening in modulus is also seen after the ODT, which is expected as the pyrochlore is generally stiffer than the fluorite [75]. At intermediate doping levels, the trend is nearly linear.

The measured thermal conductivity ($k$), as shown in Figure 7(c), also sharply changes at low concentrations of mixing on both ends, in comparison with the RoM values. However, the change is negative (with more reduced $k$ from RoM) at the P1 side, but positive at the N1 side. A more considerable drop in thermal conductivity is seen after the ODT, which may be related to lattice softening. Similar to prior discussion, possible formation of short-range chemical or structural orders may help explain the sharp decrease in thermal conductivity [76,77], at least in the P1 end. Reductions in thermal conductivity may also be due to so-called "locon" effects where certain atoms may profoundly distort the local field that allows for localized, non-propagating phonons [78–81]. The increase in thermal conductivity at the N1 end can be attributed to the increased modulus.

Notably, the $E/k$ ratios are enhanced above the RoM values consequently, compared to the two endmembers (P1 and N1, which have similar $E/k$ ratios of ~175 GPa·m·K·W$^{-1}$). This effect is more pronounced on the P1-rich side. At intermediate concentrations, the changes are typically linear. In all cases, the measured properties tend to deviate significantly from the RoM values. The ability to increase the $E/k$ ratio is scientifically interesting, as the reduction of thermal conductivity typically accompanies with decreasing modulus. In addition to potential "high-entropy" effects (including severe lattice distortion and short-range orders), potential benefits may also result with the new ability to include "non-traditional" elements (*i.e.*, elements that are not stable on their own in the phase, such as $Dy_2Zr_2O_7$ in the pyrochlore phase) in many-component solid solutions. Some of such "non-traditional" elements include Sm, Ti, Zr, and Hf in the niobate matrix and Nb in the pyrochlore matrix. Further investigation is underway to examine the influence of these "non-traditional" elements and the short-range ordering.

## 4.6 Temperature-Dependent Thermal Conductivity

Figure 8 shows the temperature-dependent thermal conductivity of all the specimens from room temperature to 1000°C. Nearly all compositions have ultralow thermal conductivity and primarily shows diffuson-like trends (with the thermal conductivity primarily rising due to heat capacity contributions). This diffuson-like behavior at high temperatures ($T > 600°C$) is similar to that of $La_2Zr_2O_7$ pyrochlore at high temperature [82]. However, at room temperature, these duodenary high-entropy oxides have much lower thermal conductivity than that of $La_2Zr_2O_7$, presumably due to the severe lattice distortion that scatters the propagating vibrational modes (propagons). It should also be noted that significant internal radiation contributions likely exist at high temperatures ($T > 600°C$), leading to a small but non-negligible photon thermal conductivity that is also increasing with temperature [65]. Qualitatively, the temperature dependence of the P1N1 series in Figure 8(a) has the same trend as the two endmembers. Quantitatively, the relative values of temperature-dependent thermal conductivity follow similar trends as the room-temperature thermal conductivity shown in Figure 7(c). However, P1N1 50-50 exhibits the lowest thermal conductivity at 1000°C, lower than the endmember N1 that has the lowest thermal conductivity at room temperature.

The temperature-dependent thermal conductivity of P1'N1' 50-50 and P1"N2 50-50 shown in Figure 8(b) has similar trends to the P1N1 series. The fluorite-structured P1"N2 50-50 has a near temperature-independent and lower thermal conductivity compared to the pyrochlore-structured P1'N1' 50-50, and it has the lowest measured thermal conductivity of <1.4 W·m$^{-1}$·K$^{-1}$ at 1000°C.

The temperature-dependent phonon and diffuson limit was computed for P1N1 50-50 and P1"N2 50-50 (the two compositions that exhibited the lowest $k$ at high temperatures) and plotted in Figure 8(a) and 8(b). The phonon limit is based on Cahill, Watson, and Pohl's seminal work in 1992 [83]:

$$k_P \approx \left(\frac{\pi}{6}\right)^{\frac{1}{3}} k_B n^{\frac{2}{3}} \sum_i v_i \left(\frac{T}{\theta_i}\right)^2 \int_0^{\frac{\theta_i}{T}} \frac{x^3 e^x}{(e^x - 1)^2} dx \qquad (6)$$

The diffuson limit is based on Allen and Feldman's work on amorphous Si [80,84,85]:

$$k_D \approx \frac{n^{-\frac{2}{3}} k_B}{2\pi^3 v_S^3} \left(\frac{k_B T}{\hbar}\right)^4 \int_0^{0.95 \frac{\theta_D}{T}} \frac{x^5 e^x}{(e^x - 1)^2} dx \quad (7)$$

Here, $n$ is the number density of atoms per unit cell, $v$ is the speed of sound, and $\theta$ is the Debye temperature. Equation (6) sums the contributions over the longitudinal and both transverse sound modes in the material while equation (7) averages these by using:

$$v_s = \left(\frac{1}{3}\left(\frac{1}{v_L} + \frac{2}{v_T}\right)\right)^{-1} \quad (8)$$

The lattice parameter, density, and longitudinal and transverse wave speeds used were determined at room temperature and were not corrected for reductions at high temperatures. As shown in Figure 8, for both specimens for which the limits were calculated for, the experimentally determined thermal conductivity approaches the phonon limit and lies within the uncertainties of the model and experiments ($\pm$ 3 %). This indicates significant contributions from diffusons, although still far from their limit. In broader perspective, the glass-like thermal conductivity of the fluorite-structured niobates has recently attracted significant research interests due to their attractive thermal properties and high temperature capabilities [25,36–38,86].

While noting the ultralow thermal conductivity and attributing the thermal conductivity primarily to diffuson contributions, the potential reasons for the amorphous-like conductivity remain elusive. A few crystallography papers dating back to the 1960s and 1970s have reported nanodomains or short-range ordering in fluorite-structured niobates and tantalates evident by diffuse scattering in electron diffraction patterns [87,88]. The nanodomain is of the weberite structure, which has an orthorhombic cell (space group 20, *C222$_1$*). The weberite structure has a close relationship to the fluorite structure (and thus pyrochlore) in that it is a ~ $\sqrt{2}a_F \times \sqrt{2}a_F \times 2a_F$ derivative of the fluorite cell [89,90]. Similar to 3+/4+ fluorites (such as zirconates) that transform to the ordered pyrochlore structure when $\frac{r_A^{VIII}}{r_B^{VI}} > 1.46$, 3+/5+ fluorites (such as niobates) transform to the partially ordered weberite structure when $\frac{r_{3+}^{VII}}{r_{5+}^{VII}} > \sim 1.40$ [91–93]. The ODT temperatures in these niobates and tantalates (e.g., Y$_3$TaO$_7$ and Ho$_3$TaO$_7$ is ~ 1500°C [88]) are lower than the pyrochlore (typically >1600°C). Thus, as our duodenary oxides were cooled after sintering at 1600°C, the precipitation of nanoscale weberite domains (a.k.a. short-range structural ordering) is possible. The strong effect of dispersed nanodomains on reducing thermal conductivity [77,94] can be a possible cause of the ultralow, diffuson-like thermal conductivity observed. Further (non-trivial) experiments and characterization are needed to verify or clarify the exact underlying mechanism.

# 5 Conclusions

A five-cation fluorite-structured high-entropy niobate and a seven-cation high-entropy pyrochlore were mixed in different fractions and consolidated to make a series of duodenary oxides. All seven (plus two additional) compositions made are single phase high-entropy solid solutions, where an order-disorder transition (ODT) is evident with changing composition. The critical ODT composition in these duodenary oxides is shifted with respect to that predicted based on two criteria proposed based on ternary oxides. These duodenary oxides are more prone to disorder, which are attributed to a "high-entropy" effect and the Sc deficiency on the B-site.

Increases in measured Young's modulus above the rule-of-mixture values are observed, with abrupt enhancements at low mixing compositions near the two endmembers. Consequently, the $E/k$ ratios of the duodenary oxides are higher than those of both the endmembers. Nearly all of the compositions show glass-like temperature-dependent thermal conductivity. Possible existence and effects of short-range ordering (forming nanodomains) in these complex duodenary oxides are discussed.

This study represents the first report of a series of duodenary HECs and the first investigation an ODT in HECs with changing composition and its influence on thermomechanical properties. It further suggests a new route to tailor the properties of HECs via controlling the order and disorder.

**Acknowledgment:** This work is currently supported by the U.S. National Science Foundation (NSF) via Grant No. DMR-2026193 in the Ceramics program. We also acknowledge a seed project (Small Innovative Projects in Solar) supported by the Solar Energy Technologies Office in the U.S. Department of Energy's Office of Energy Efficiency and Renewable Energy under Contract No. EE0008529, where we initiated this work before the NSF project started. The STEM work was performed at the Irvine Materials Research Institute (IMRI).

**Table 1.** Compositions examined in this study. All specimens are made by mixing different fractions of the pyrochlore (P1, P1' and P1") and niobate (N1, N1' and N2) endmembers of nominally 0% mismatches in both lattice and density. Here, P1', P1" and N1' are derivative compositions of P1 and N1 with varying atomic ratios, and "$x$" denotes the fraction of the fluorite phase before the mixing (for the P1N1 series). For example, "P1N1 10-90" (*i.e.*, $x = 0.9$ in the P1N1 series in this table) in text represents the composition starting from mixing 10 % of the P1 pyrochlore and 90 % of the N1 niobate (albeit it exhibits a single-phase fluorite structure after mixing). In fact, all nine mixed compositions are single-phase (either fluorite or pyrochlore) high-entropy solid solutions after mixing.

| | | Compositions ($x$ = 0, 0.02, 0.1, 0.25, 0.5, 0.75, 0.9, 0.98, 1) |
|---|---|---|
| P1N1 Series | P1 | $[(Sm_{0.25}Eu_{0.25}Gd_{0.25}Yb_{0.25})_2(Ti_{0.5}Hf_{0.25}Zr_{0.25})_2O_7]_{1-x}$ |
| | N1 | $[(Sc_{0.266}Dy_{0.248}Tm_{0.246}Yb_{0.240})_3NbO_7]_x$ |
| P1'N1' 50-50 | P1' | $[(Sm_{0.25}Eu_{0.25}Gd_{0.25}Yb_{0.25})_2(Ti_{0.611}Hf_{0.166}Zr_{0.223})_2O_7]_{0.5}$ |
| | N1' | $[(Sc_{0.333}Dy_{0.226}Tm_{0.224}Yb_{0.218})_3NbO_7]_{0.5}$ |
| P1"N2 50-50 | P1" | $[(Sm_{0.25}Eu_{0.25}Gd_{0.25}Yb_{0.25})_2(Ti_{0.065}Hf_{0.234}Zr_{0.701})_2O_7]_{0.5}$ |
| | N2 | $[(Y_{0.266}Dy_{0.248}Tm_{0.246}Yb_{0.240})_3NbO_7]_{0.5}$ |

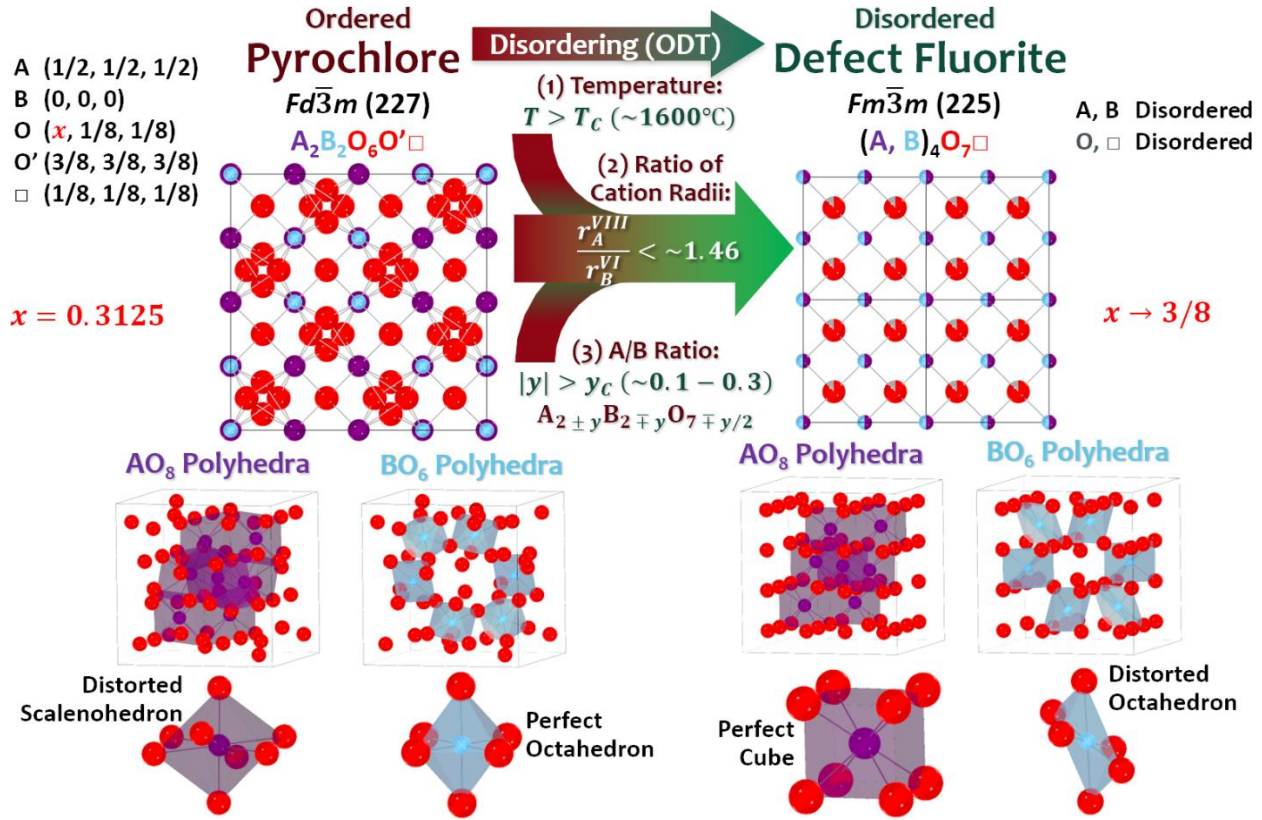

**Figure 1.** Schematic illustration of the ordered pyrochlore and disordered defect fluorite structures and the associated order-disorder transition (ODT). In simple ternary $A_2B_2O_7$ pyrochlore oxides, three criteria to induce an ODT are through (1) increasing temperature ($T > T_c$), (2) reducing the ratio of cation radii ($r_A^{VIII}/r_B^{VI} > \sim 1.46$, where $r_A^{VIII}$ is the average radius of the A-site cation with a coordination number of eight and $r_B^{VI}$ is the average radius of the B-site cations with a coordination number of six), and (3) increasing the non-stoichiometry ($|y| > y_c$, *i.e.*, deviating from the ideal 1:1 A/B ratio). In the ideal ordered $A_2B_2O_7$ (or $A_2B_2O_6O'\square$) pyrochlore structure, there are 1/8 ordered oxygen vacancies (represented by □), and six oxygen atoms are displaced to ($x$, 1/8, 1/8), where $x = 0.3125$. Upon an ODT, oxygen atoms are at the ideal (3/8, 1/8, 1/8) or $x = 3/8$ in the fluorite structure, accompanying with disordering on both cation (A and B) and anion (oxygen ions and vacancies) sublattices. Note that the $AO_8$ and $BO_6$ polyhedrons are shown to represent the transformation merely; in real disordered fluorite structure, however, cations are randomly distributed on the same site to form $(A, B)O_7$ polyhedrons with an average coordination number of VII (instead of perfect $AO_8$ and $BO_6$ polyhedrons).

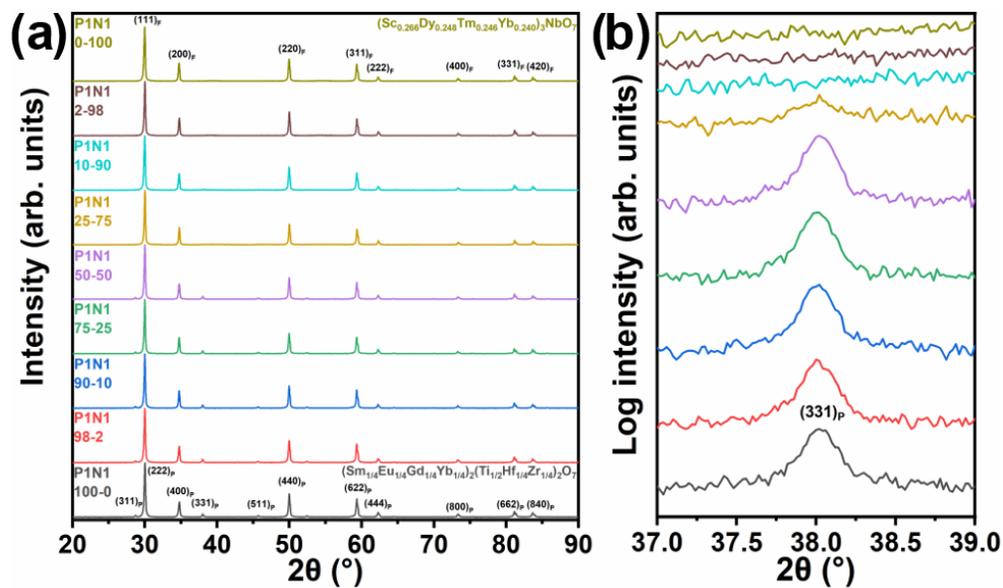

**Figure 2.** (**a**) XRD spectra of P1N1 series and (**b**) the evolution of (331) pyrochlore superstructure peak. The evolution of the pyrochlore superstructure peak is plotted on a logarithmic intensity scale.

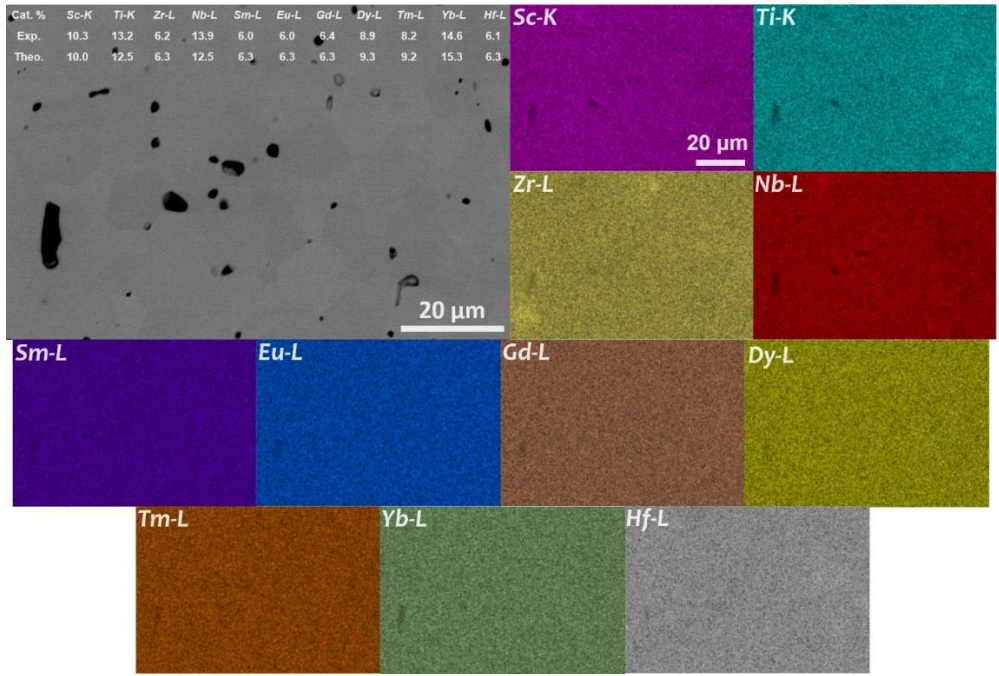

**Figure 3.** Backscattered electron SEM microstructure and EDS elemental maps of P1N1 50-50. The composition is predominantly homogeneous, albeit slight Zr agglomeration. The grain size is about 10 – 20 μm. The nominal (theoretical) and measured metal cation percentages are listed in the SEM micrograph, which agree well with each other within the experimental errors. Additional EDS elemental maps are documented in Suppl. Figs. S2 and S3, showing homogeneous compositions before and after the ODT.

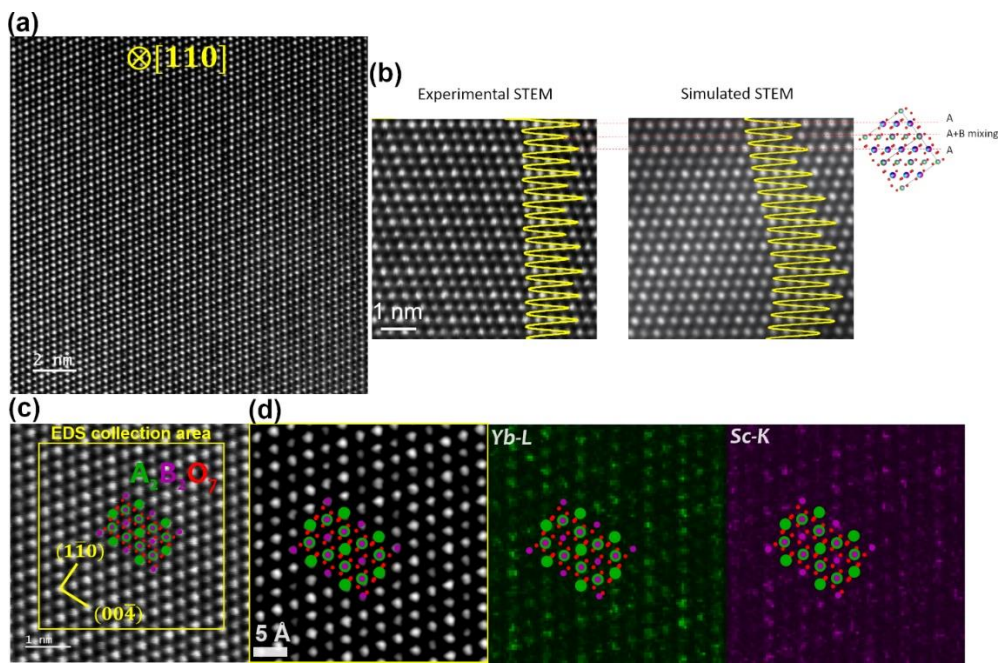

**Figure 4.** Raw (unfiltered) STEM-HAADF images of P1N1 50-50 at (**a**) low and (**b**) high magnification viewed along the [110] zone axis, showing intensity modulations due to the cation ordering. (**c**) Atomic resolution STEM was performed on a small area. (**d**) Bandpass filtered STEM-HAADF image and radial Wiener filtered EDS elemental maps of Yb (on the A site) and Sc (on the B site).

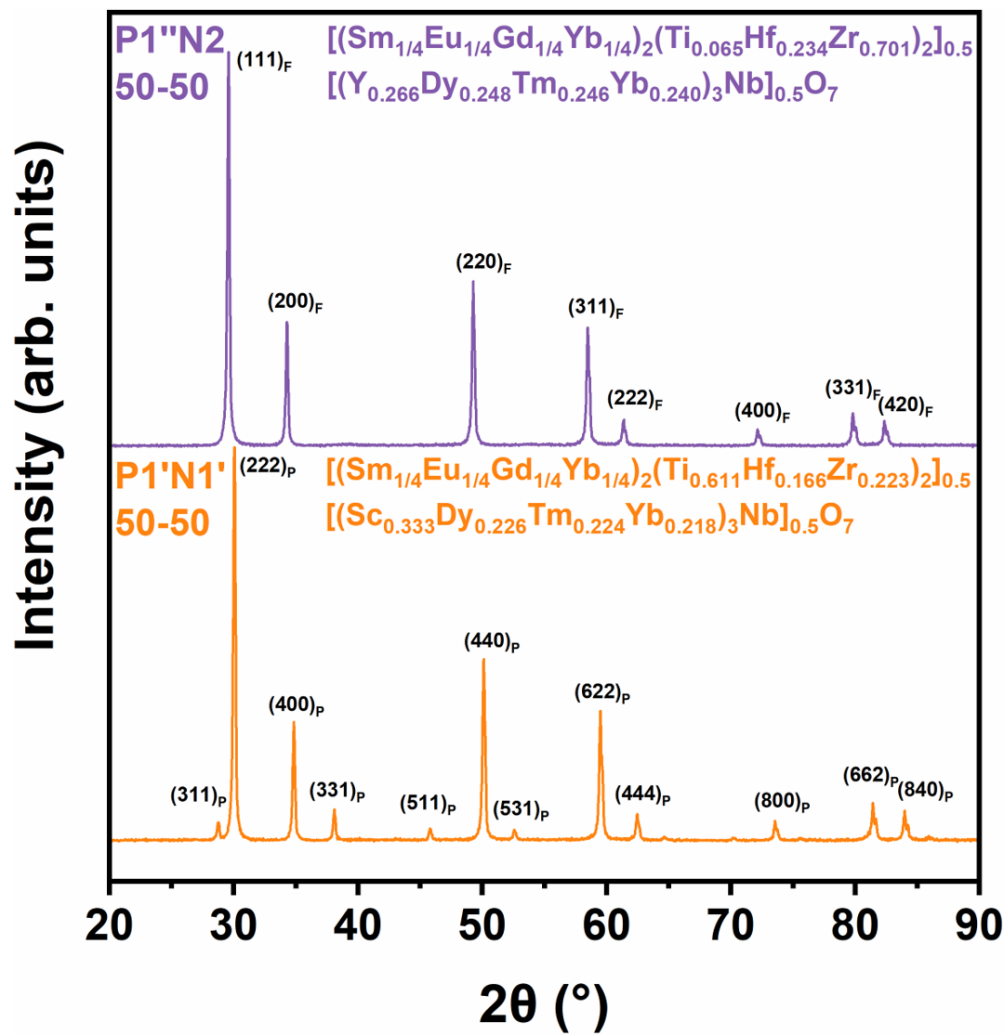

**Figure 5.** XRD spectra of the pyrochlore-structured P1'N1' 50-50 (containing Sc) and the fluorite-structured P1"N2 50-50 (without Sc).

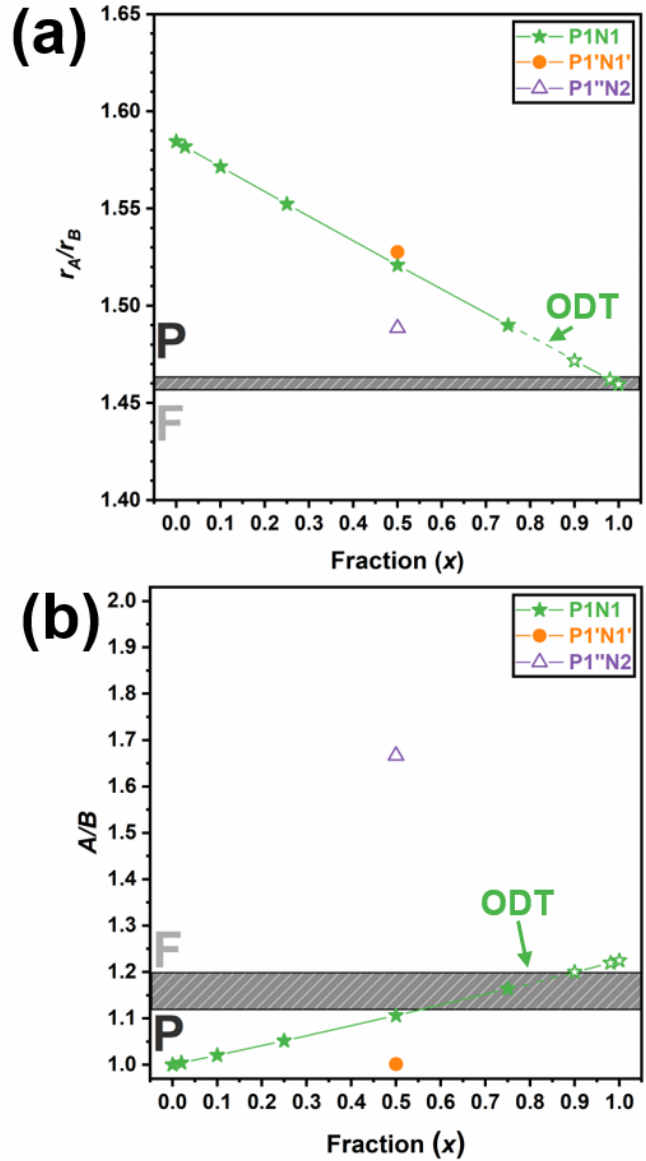

**Figure 6.** Key factors controlling the relative stabilities of the disordered fluorite (F) *vs.* ordered pyrochlore (P). The estimated (**a**) ratio of cation radii ($r_A/r_B$) and (**b**) stoichiometric ratio of A/B cations. The striped region marks the expected pyrochlore-to-fluorite transition from ternary oxides (with the thickness representing errors); see text for further explanation. Filled and open data points represent the high-entropy pyrochlore and fluorite phases, respectively, from this study, which do not exactly follow the empirical laws primarily obtained from ternary oxides.

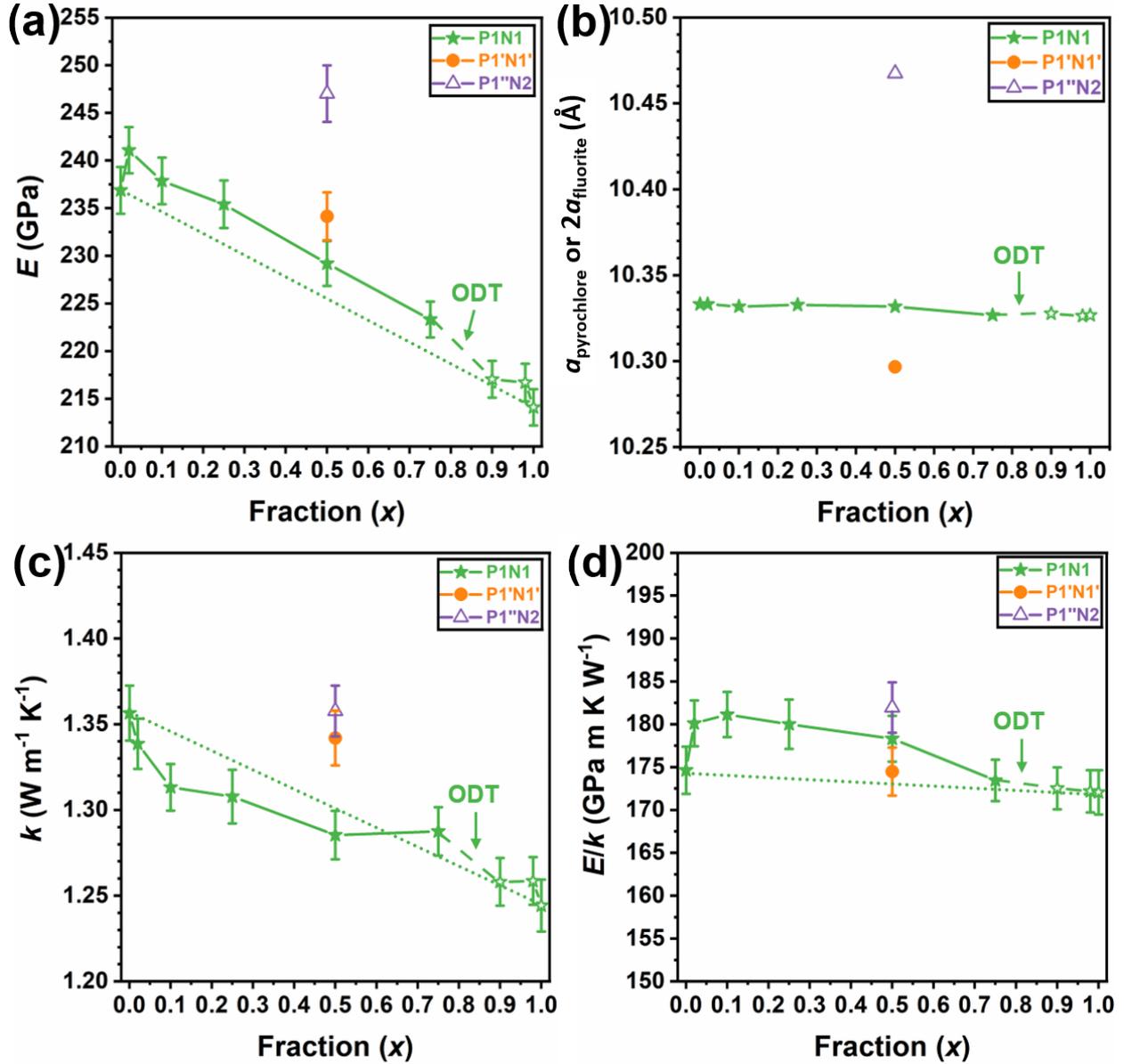

**Figure 7.** Room-temperature (**a**) Young's modulus ($E$), (**b**) lattice parameter, (**c**) thermal conductivity ($k$), and (**d**) $E/k$ measured for the P1N1 series and two single specimens, P1'N1' 50-50 and P1"N2 50-50. Arrows and dashed lines are used to denote the order-to-disorder (pyrochlore-to-fluorite) transition (ODT). Filled and open data points represent the pyrochlore and fluorite phases, respectively. The dotted lines represent the rule-of-mixture averages from two endmembers for the P1N1 series.

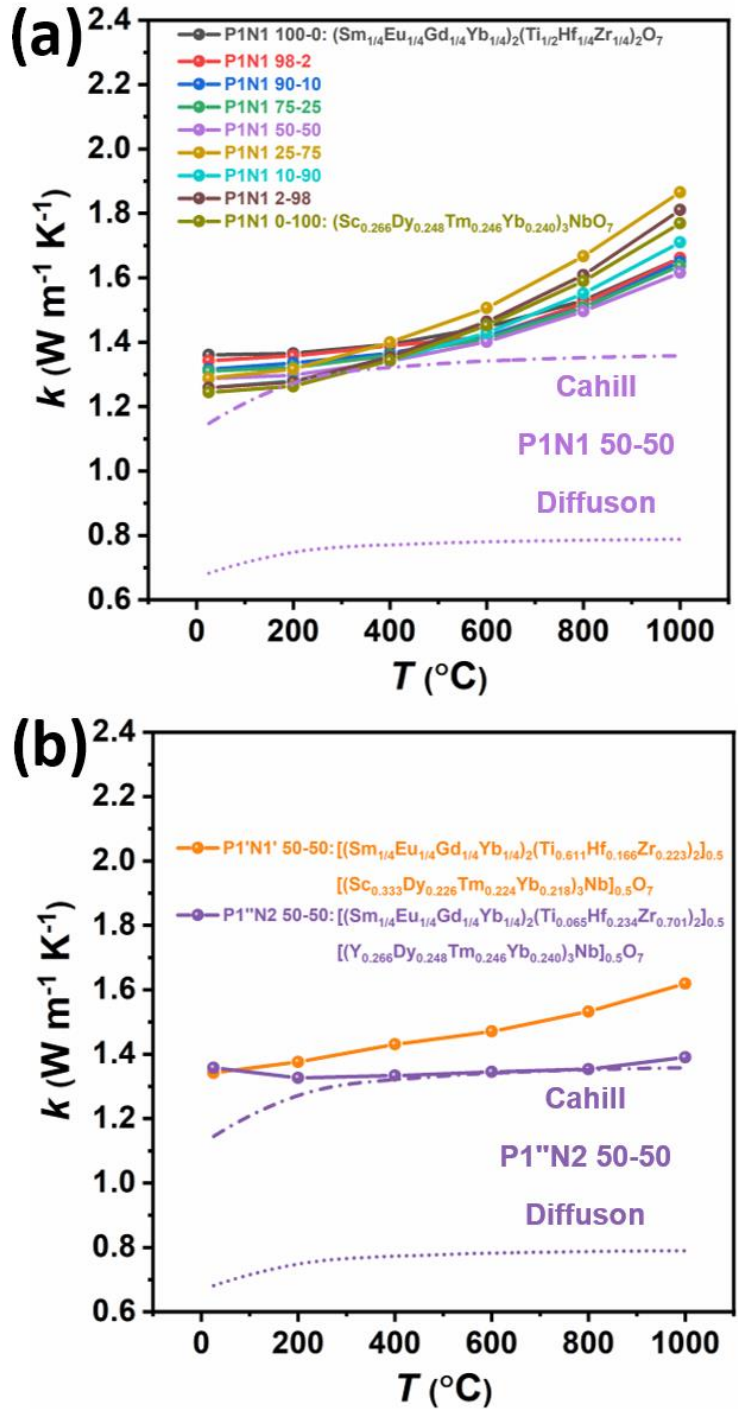

**Figure 8.** Temperature-dependent thermal conductivity of (**a**) P1N1 and (**b**) P1'N1' 50-50 and P1"N2 50-50. Cahill and diffuson model limits are shown for P1N1 50-50 and P1"N2 50-50, respectively (representing the compositions with the lowest *k* at high temperatures in each panel).